\documentclass[preprint,12pt]{elsarticle}
\usepackage[utf8]{inputenc}
\usepackage{amssymb}
\usepackage{graphicx}
\usepackage{multirow}%
\usepackage{amsmath,amssymb,amsfonts,bm}%
\usepackage{amsthm}%
\usepackage{multirow}
\usepackage{mathrsfs}%
\usepackage[title]{appendix}%
\usepackage{float}
\usepackage{enumitem}
\usepackage{placeins}
\usepackage{epsfig}
\usepackage{xcolor}%
\usepackage{textcomp}%
\usepackage{manyfoot}%
\usepackage{booktabs}%
\usepackage{enumitem}
\usepackage{array}
\usepackage{geometry}
\usepackage{subcaption}
\usepackage{algorithm}%
\usepackage{algorithmicx}%
\usepackage{algpseudocode}%
\usepackage{listings}%
\usepackage{dashrule}
\usepackage{hyperref}

\hypersetup{
    colorlinks=true,
    linkcolor=blue,
    citecolor=red,
    urlcolor=blue
}
\journal{}

\begin{document}

\begin{frontmatter}

\title{An Exponential-Polynomial Divergence-based Robust Information Criterion for Linear Panel Data Models and Neural Networks}

\author[inst1]{Udita Goswami}

\affiliation[inst1]{organization={Department of Mathematics and Computing},
            addressline={IIT(Indian School of Mines) Dhanbad}, 
            city={},
            postcode={826004}, 
            state={Jharkhand},
            country={India}}

\author[inst1]{Shuvashree Mondal*}

\begin{abstract}

Model selection is a cornerstone of statistical inference, where information criteria are widely employed to balance model fit and complexity.  However, classical likelihood-based criteria are often highly sensitive to contamination, outliers, and model misspecification.  In this paper, we develop a robust alternative based on the Exponential-Polynomial Divergence, a flexible extension of existing divergence measures that enhances adaptability to diverse data irregularities.  The proposed Exponential-Polynomial Divergence Information Criterion preserves the objective of approximating the discrepancy between the true model and candidate models while incorporating robustness against anomalous observations.  Its theoretical properties are established, and robustness is examined through influence function analysis, demonstrating controlled sensitivity to extreme data points.  For practical implementation, a data-driven tuning parameter selection strategy based on generalized score matching is employed, ensuring improved computational stability and efficiency.  The effectiveness of the proposed method is demonstrated through extensive simulation studies under varying contamination levels, as well as real data applications involving linear mixed-effects panel data models and neural network-based prediction tasks.  The results consistently show improved stability and reliability compared to classical likelihood and density power divergence-based information criteria. The proposed framework thus provides a practical and unified approach for model selection in complex and contaminated data settings.

\end{abstract}


\end{frontmatter}

\section{Introduction}\label{sec1}

Information criteria form a fundamental class of tools for model selection, balancing goodness-of-fit and model complexity within an information-theoretic framework.  Their construction is rooted in likelihood-based inference, where the objective is to maximize the log-likelihood while accounting for model complexity, leading to an approximation of the expected Kullback–Leibler divergence between the true model and a candidate model. The Akaike Information Criterion (AIC), proposed by Hirotugu Akaike, provides an approximately unbiased estimator of this divergence under correct model specification, whereas the Takeuchi Information Criterion (TIC) extends this idea to settings where the model may be misspecified \citep{akaike1998information, takeuchi1976distribution}.  Recent work further shows that, for non-normalized models where the likelihood is intractable, information criteria can still be formulated as approximately unbiased estimators of suitable discrepancy measures, thereby enabling principled model selection beyond the classical likelihood framework \citep{matsuda2021information}.

Despite their widespread applicability and theoretical appeal, classical information criteria are often unsuitable in the presence of contamination, outliers, or model misspecification, as they rely on likelihood-based objective functions that are inherently non-robust.  Even a small fraction of aberrant observations can significantly distort parameter estimates and, consequently, the model selection outcome.  As a result, criteria such as AIC may lead to unreliable conclusions when the assumed model deviates from reality.  This limitation is particularly relevant in practice, where data from socio-economic systems, industrial processes, healthcare, and environmental studies are frequently affected by noise, anomalies, and departures from ideal assumptions, highlighting the need for more robust alternatives.

In an effort to address the lack of robustness in likelihood-based model selection, \citep{mattheou2009model} introduced a divergence-based information criterion (DIC) based on a generalized measure of statistical discrepancy, closely linked to the Density Power Divergence (DPD) framework of \citep{basu1998robust}.  The DPD and its subsequent developments \citep{ghosh2018new, ghosh2020ultrahigh, ray2022characterizing} provide a robust alternative to likelihood by downweighting the influence of atypical observations.  While this marks a clear departure from classical likelihood-based approaches, the practical utility of such divergence-based criteria remains underexplored.  Existing studies on DIC, including improvements and applications such as \citep{mantalos2010improved}, are largely confined to controlled simulations, with limited validation on real-world datasets.  This reveals a gap in understanding their empirical performance and highlights the need for more practically viable and robust model selection frameworks.

Motivated by the growing emphasis on robust statistical methodologies, we propose a novel information criterion grounded in the Exponential-Polynomial Divergence (EPD), a flexible divergence family introduced by \citep{singh2021robust}.  The EPD extends the Density Power Divergence by incorporating an exponential-polynomial structure, allowing greater flexibility in handling diverse contamination patterns and improving robustness without significant loss of efficiency, as also demonstrated in recent applications \citep{kim2024robust}.  Building upon this framework, we propose the Exponential-Polynomial Divergence Information Criterion (EPDIC), which leverages the robustness of EPD to enable reliable model selection under contamination and model misspecification, providing a practical alternative to existing likelihood- and DPD-based criteria.

A key contribution of this work is the assessment of the robustness of the proposed EPDIC using the influence function, a standard tool for measuring sensitivity to contamination.  By analyzing its influence function, we characterize the local robustness of EPDIC and its resistance to outliers.  In particular, the boundedness of the influence function demonstrates that the proposed criterion effectively controls the impact of extreme observations, thereby inheriting strong robustness properties from the underlying divergence structure and ensuring stability under contamination.

The choice of optimal tuning parameters is crucial in divergence-based methods, as it governs the trade-off between robustness and statistical efficiency.  Classical approaches, such as the Warwick and Jones criterion \citep{warwick2005choosing} and its iterative extension \citep{basak2021optimal}, are widely used but often suffer from computational burden and instability, particularly in complex or high-dimensional settings.  Recent studies \citep{goswami2025inequality} indicate that score-based approaches can provide improved and more reliable performance.  Motivated by this, we adopt the generalized score matching (GSM) framework of \citep{xu2025generalized}, which builds on the original method of \citep{hyvarinen2005estimation}.  By avoiding the need for normalization constants and relying on derivatives of log-densities, this approach offers a computationally efficient and stable mechanism for obtaining data-adaptive tuning parameter estimates.

To substantiate the practical relevance of the proposed methodology, we conduct an extensive empirical investigation encompassing both controlled simulations and real-world data applications, which form the principal crux of this study.  The simulation framework is carefully designed to assess the finite-sample performance of the proposed exponential-polynomial divergence information criterion across varying levels of contamination, thereby providing a comprehensive understanding of its robustness.  Beyond simulations, we examine the efficacy of EPDIC in the context of linear mixed-effects panel data models, where the underlying objective function is constructed via the density power divergence measure to accommodate unobserved heterogeneity and potential contamination.  In parallel, we explore its applicability in modern machine learning settings, particularly in neural network-based prediction tasks, where the conventional loss functions are replaced or augmented by divergence-based counterparts to enhance robustness against noisy and corrupted inputs.  These applications are novel in the sense that the performance of robust information criteria has seldom been systematically evaluated across such diverse real-world scenarios.  In both simulation and empirical analyses, we provide a thorough comparative assessment of EPDIC against classical likelihood-based criteria and existing divergence-based alternatives, including those derived from the density power divergence, across multiple contamination regimes.  The results consistently demonstrate the superior stability and reliability of the proposed criterion in challenging data environments.

The remainder of this paper is organized as follows.  In Section \ref{sec2}, we rigorously introduce the Exponential-Polynomial Divergence estimator.  Section \ref{sec3} presents the formulation of the EPDIC along with its theoretical properties.  In Section \ref{sec4}, we investigate the robustness characteristics of the proposed criterion through an influence function analysis.  Section \ref{sec5} is devoted to the selection of optimal tuning parameters using the GSM approach.  Section \ref{sec6} provides an extensive empirical evaluation, including both simulation studies and real data applications.  Finally, Section \ref{sec7} concludes the paper with a summary of findings and potential directions for future research.

\section{Exponential-Polynomial Divergence  Estimator}\label{sec2}
\allowdisplaybreaks

The Exponential-Polynomial Divergence (EPD), proposed by \citep{singh2021robust}, represents a unified and generalized class of Bregman divergences, capable of encompassing several well-known divergence measures such as the Density Power Divergence (DPD), Bregman Exponential Divergence (BED), and Kullback-Leibler (KL) divergence as special cases.

Under the assumptions originally stated by \citep{bregman1967relaxation}, the Bregman divergence between two density functions $g$ and $f$ is defined as
\begin{equation}
    D_B(g, f) = \int_x \left[B(g(x)) - B(f(x)) - \{g(x) -f(x)\}B'(f(x))\right]dx,
\label{eq1}
\end{equation}
where $B(\cdot)$ is a strictly convex function and $B'(\cdot)$ denotes its derivative with respect to its argument.  The choice of the convex generating function $B$ determines the specific form of the divergence; different selections of $B$ lead to distinct members within the Bregman divergence family (see, e.g., \citep{banerjee2005clustering}, \citep{cichocki2010families}). 

To generalize the existing divergence families, \citep{singh2021robust} introduced a convex generating function of the form 
\begin{equation}
B(x) = \frac{\beta}{\alpha^2}\left(e^{\alpha x} - 1 - \alpha x\right) + \frac{1 - \beta}{\gamma}
\left(x^{\gamma + 1} - x\right), \quad
\alpha \in \mathbb{R},\; \beta \in [0, 1],\; \gamma \ge 0,
\label{eq2}
\end{equation}
where $\alpha$, $\beta$, and $\gamma$ are tuning parameters that control the contributions of the exponential and polynomial components.  The EPD connects several well-known divergences through particular parameter choices:
\begin{itemize}
    \item When $\beta = 0$, the EPD reduces to the Density Power Divergence (DPD) of \citep{basu1998robust} with parameter $\gamma$.
    \item When $\beta = 1$, it coincides with the Bregman Exponential Divergence (BED) of \citep{nielsen2013chi} with parameter $\alpha$.
    \item For intermediate values $0<\beta<1$, it provides a convex combination of BED and DPD.
    \item Further, when $\beta = 0$ and $\gamma \to 0$, the divergence converges to the Kullback-Leibler divergence \citep{kullback1951information}.
\end{itemize}

\noindent Thus, the EPD offers a smooth continuum of divergence measures linking the exponential-type and power-type divergences through its three-parameter structure.

\subsection{\textbf{Estimation under Independent Non-homogeneous observations}}
\allowdisplaybreaks

Consider independent but not identically distributed observations $Y_1, Y_2, \ldots, Y_n$, where each $Y_i$ follows a potentially different density $g_i$ are modeled by a parametric family $\mathcal{F}_{i,\theta} = \left\{ f_i(\cdot; \theta) \,\middle|\, \theta \in \Theta \subseteq \mathbb{R}^p \right\},$ sharing the same common parameter $\theta$.  The goal is to estimate $\theta$ robustly by minimizing the average exponential-polynomial divergence between the true and model densities, 
\begin{align}
\frac{1}{n}\sum_{i=1}^n D_{\mathrm{EP}}\big(g_i, f_i(\cdot; \theta)\big) 
&= \frac{1}{n}\sum_{i=1}^n \Bigg[
\int \Big\{ B'\big(f_i(y; \theta)\big) f_i(y; \theta) - B\big(f_i(y; \theta)\big) \Big\} dy \\ \nonumber
&\quad - B'\big(f_i(Y_i; \theta)\big)
\Bigg].
\label{eq3}
\end{align}

\noindent where $D_{\text{EP}}(\cdot, \cdot)$ denotes the divergence corresponding to the generating function $B(\cdot)$ in \eqref{eq2}.  

Since only a single observation $Y_i$ is available from each density $g_i$, we approximate $g_i$ by the degenerate empirical distribution that places unit mass at $Y_i$.  The resulting empirical objective function for estimation becomes
\begin{equation}
H_n^{(\alpha, \beta, \gamma)}(\theta)
= \frac{1}{n}\sum_{i=1}^n
V_{\alpha, \beta, \gamma}(Y_i; \theta),
\label{eq4}
\end{equation}
where $V_{\alpha, \beta, \gamma}(Y_i; \theta)$ represents the samplewise contribution to the divergence and combines exponential and polynomial components as
\begin{align}
V_{\alpha, \beta, \gamma}(Y_i; \theta)
&=
\frac{\beta}{\alpha^2} \int_y \left[ \{e^{\alpha f_i(y; \theta)}(\alpha f_i(y; \theta) - 1) + 1\} dy + (1-\beta)f^{1+\gamma}_i(y; \theta)\right] dy
\nonumber \\[6pt]
&\quad -
\left[ \frac{\beta}{\alpha} (e^{\alpha f_i(Y_i; \theta)} - 1) + \frac{1 - \beta}{\gamma} ((\gamma+1)f_i(Y_i; \theta) - 1)
\right].
\label{eq5}
\end{align}

\noindent Minimizing $H_n^{(\alpha, \beta, \gamma)}(\theta)$ with respect to $\theta$ yields the Minimum Exponential-Polynomial Divergence Estimator (MEPDE).  Now, when we differentiate $H_n^{(\alpha, \beta, \gamma)}(\theta)$ with respect to $\theta$, we get the following estimating equation:
\begin{equation}
\frac{1}{n}\sum_{i=1}^n
\left[
u_i(f_i(Y_i; \theta))\, B''_i(Y_i; \theta) f_i(Y_i; \theta)
-
\int_y u_i(f_i(y; \theta))\, B''_i(y; \theta) f^{2}_i(y; \theta)\, dy
\right] = 0,
\label{eq6}
\end{equation}
where $u_i(y; \theta) = \frac {\partial}{\partial \theta} \log f_i(y; \theta)$ is the likelihood score function of the $i^{th}$ sample.  Further, we can express the proposed formulation as a weighted likelihood estimating equation, viz.,
\begin{equation}
\frac{1}{n} \sum_{i=1}^n
\left[
u_i(f_i(Y_i; \theta))\, w(f_i(Y_i; \theta))
-
\int_y u_i(f_i(y; \theta))\, w(f_i(y; \theta)) f_i(y; \theta)\, dy
\right] = 0,
\label{eq7}
\end{equation}
where the corresponding weight function is of the form
\[
w(t) = \beta t e^{\alpha t} + (1-\beta)(1+\gamma) t^{\gamma}.
\]
\noindent The above estimating equation generalizes the minimum density power divergence estimator (MDPDE) for non-homogeneous observations proposed by \citep{ghosh2013robust}, while preserving the bias-variance trade-off controlled by $(\alpha, \beta, \gamma)$.  When $\beta = 0$, the MEPDE reduces to the MDPDE, and as $(\beta, \gamma, \alpha) \to (0, 0, 0)$, it simplifies to the maximum likelihood estimating equation:
\begin{equation}
\frac{1}{n} \sum_{i=1}^n u_i(Y_i; \theta) = 0.
\label{eq8}
\end{equation}

\noindent From a functional perspective, the minimum Exponential-Polynomial divergence functional for independent non-homogeneous data is defined as
\begin{equation}
T_{\alpha,\beta,\gamma}(G_1, \ldots, G_n)
= \arg\min_{\theta \in \Theta}
\frac{1}{n}\sum_{i=1}^n
D_{\text{EP}}(g_i, f_i(\cdot; \theta)).
\label{eq9}
\end{equation}
Since $D_{\text{EP}}(\cdot, \cdot)$ is a valid divergence---non-negative and equal to zero if and only if $g_i = f_i(\cdot; \theta)$---the functional $T_{\alpha,\beta,\gamma}(G_1, \ldots, G_n)$ is Fisher consistent under model identifiability, satisfying
\begin{equation}
T_{\alpha,\beta,\gamma}(F_{1,\theta_0}, \ldots, F_{n,\theta_0}) = \theta_0.
\label{eq10}
\end{equation}
Hence, the proposed MEPDE serves as a natural and robust generalization of the MDPDE for independently but non-identically distributed data, combining the strengths of both exponential and polynomial divergence families within a unified estimation framework.

\subsection{\textbf{Asymptotic Properties}}
\allowdisplaybreaks

To study the asymptotic behavior of the Minimum Exponential-Polynomial Divergence Estimator (MEPDE) under independent non-homogeneous observations, let
\[
\theta_g = \arg\min_{\theta \in \Theta}\, \frac{1}{n}\sum_{i=1}^n D_{\text{EP}}(g_i, f_i(\cdot; \theta))
\]
denote the best-fitting parameter that minimizes the average exponential-polynomial divergence between the true and model densities.  Let us introduce a $p\times p$ matrix denoted by $\boldsymbol{J}_i$, whose $(k,l)^{th}$ entry is defined through the second-order partial derivative with respect to the $k^{th}$ and $l^{th}$ components of $\theta$.  In addition, we define the quantities $\boldsymbol{K}_i$ and $\boldsymbol{\xi}_i$ as given below.  

\begin{align}
\boldsymbol{J}_i &= 
\beta \int_y f_i^2(y; \theta_g)\,
    e^{\alpha f_i(y; \theta_g)} 
    u_i(y; \theta_g)u_i^{\top}(y; \theta_g)\, dy \notag \\[4pt]
&\quad + (1-\beta)(1+\gamma) 
    \int_y f_i^{\gamma+1}(y; \theta_g)\,
    u_i(y; \theta_g) u_i^{\top}(y; \theta_g)\, dy \notag \\[4pt]
&\quad + (1-\beta)(1+\gamma)
    \int_y \!\left[g_i(y) - f_i(y; \theta_g)\right]
    \left\{\boldsymbol{I}_i(y, \theta_g) - 
    \gamma\, u_i(y; \theta_g) u_i^{\top}(y; \theta_g)\right\}
    f_i^{\gamma}(y; \theta_g)\, dy \notag \\[4pt]
&\quad + \beta \int_y \!\left[g_i(y) - f_i(y; \theta_g)\right]
    \left\{\boldsymbol{I}_i(y, \theta_g) - 
    u_i(y; \theta_g) u_i^{\top}(y; \theta_g)\right\}
    f_i(y; \theta_g)e^{\alpha f_i(y; \theta_g)}\, dy \notag \\[4pt]
&\quad - \alpha \beta 
    \int_y \!\left[g_i(y) - f_i(y; \theta_g)\right]
    f_i^2(y; \theta_g)e^{\alpha f_i(y; \theta_g)}
    u_i(y; \theta_g) u_i^{\top}(y; \theta_g)\, dy.
\label{eq11}
\end{align}

\begin{align}
\boldsymbol{K}_i 
&= 
\int_y 
u_i(y; \theta_g)\, u_i^{\top}(y; \theta_g)
\Big\{
\beta\, f_i(y; \theta_g) e^{\alpha f_i(y; \theta_g)}
+ (1 - \beta)(1 + \gamma)\, f_i^{\gamma}(y; \theta_g)
\Big\} g_i(y)\, dy
- \boldsymbol{\xi}_i \boldsymbol{\xi}_i^{\top}.
\label{eq12}
\end{align}
where
\begin{align}
\boldsymbol{\xi}_i 
&= \int_y 
u_i(y; \theta_g)
\Big\{
\beta f_i(y; \theta_g)e^{\alpha f_i(y; \theta_g)}
+ (1-\beta)(1+\gamma)f_i^{\gamma}(y; \theta_g)
\Big\}
g_i(y)\, dy.
\label{eq13}
\end{align}

\medskip
\noindent \textbf{Regularity Conditions.}\\

\noindent To establish consistency and asymptotic normality, we assume conditions analogous to those in \citep{ghosh2013robust} and \citep{singh2021robust}:

\begin{enumerate}[label=(R\arabic*)]
    \item The support $\mathcal{X} = \{y : f_i(y; \theta) > 0\}$ is independent of both $i$ and $\theta$, for all $i = 1, 2, \ldots, n$, and the true densities $g_i$ are also supported on $\mathcal{X}$.

    \item There exists an open subset $\Theta_0 \subseteq \Theta$ containing the best-fitting parameter $\theta_g$ such that, for almost all $y \in \mathcal{X}$ and all $\theta \in \Theta_0$, the densities $f_i(y; \theta)$, $i = 1, 2, \ldots, n$, are three times continuously differentiable with respect to $\theta$, and all third-order partial derivatives are continuous in $\theta$.

    \item For each $i = 1, 2, \ldots, n$, and for the generating function $B(\cdot)$ defined in \eqref{eq2}, the integrals
    \[
    \int_y B'(f_i(y; \theta)) f_i(y; \theta) \, dy
    \quad \text{and} \quad
    \int_y B'(f_i(y; \theta)) g_i(y) \, dy
    \]
    can be differentiated three times with respect to $\theta$, and differentiation can be interchanged with integration.

    \item For each $i = 1, 2, \ldots, n$, the matrix $\boldsymbol{J}_i$ defined in \eqref{eq11} is positive definite. 
    We define the average information matrix
    \[
    \boldsymbol{\Psi}_n = \frac{1}{n}\sum_{i=1}^{n} \boldsymbol{J}_i .
    \]
    Assume that the sequence $\{\boldsymbol{\Psi}_n\}$ converges to a positive definite limit
    \[
    \boldsymbol{\Psi} = \lim_{n\to\infty} \boldsymbol{\Psi}_n ,
    \]
    and that 
    \[
    \lambda_0 = \lambda_{\min}(\boldsymbol{\Psi}) > 0 .
    \]

    \item There exist measurable bounding functions $M^{(i)}_{jkl}(y)$ such that, for all $\theta \in \Theta_0$,
    \[
    \Bigg|\frac{\partial}{\partial \theta_j \partial \theta_k \partial \theta_l} V_{\alpha,\beta,\gamma}(y; \theta)\Bigg| \le M^{(i)}_{jkl}(y),
    \]
    where $V_{\alpha,\beta,\gamma}(\cdot; \theta)$ is defined in \eqref{eq5}, and
    \[
    \frac{1}{n}\sum_{i=1}^n E_{g_i}\!\left[M^{(i)}_{jkl}(Y_i)\right] = O(1)
    \quad \text{for all } j, k, l.
    \]

    \item For all $j$ and $k$, the following uniform integrability conditions hold for the first and second derivatives of $V_{\alpha,\beta,\gamma}(Y_i; \theta)$:
    \begin{align*}
    &\lim_{n \to \infty} \sup_n 
    \frac{1}{n}\sum_{i=1}^n 
    E_{g_i}\!\left[
    \Bigg|\frac{\partial}{\partial \theta_j} V_{\alpha,\beta,\gamma}(Y_i; \theta)\Bigg| \,
    \mathbb{I}\!\left\{
    \Bigg|\frac{\partial}{\partial \theta_j} V_{\alpha,\beta,\gamma}(Y_i; \theta)\Bigg|
    > \epsilon \sqrt{n}
    \right\}
    \right] = 0,
    \\[10pt]
    &\lim_{n \to \infty} \sup_n 
    \frac{1}{n}\sum_{i=1}^n 
    E_{g_i}\!\left[
    \begin{aligned}-
    &\Bigg|
    \frac{\partial}{\partial \theta_j \partial \theta_k}
    V_{\alpha,\beta,\gamma}(Y_i; \theta)
    -
    E_{g_i}\!\Bigg[
    \frac{\partial}{\partial \theta_j \partial \theta_k}
    V_{\alpha,\beta,\gamma}(Y_i; \theta)
    \Bigg]
    \Bigg|
    \\
    &\times
    \mathbb{I}\!\left\{
    \Bigg|
    \frac{\partial}{\partial \theta_j \partial \theta_k}
    V_{\alpha,\beta,\gamma}(Y_i; \theta)
    -
    E_{g_i}\!\Bigg[
    \frac{\partial}{\partial \theta_j \partial \theta_k}
    V_{\alpha,\beta,\gamma}(Y_i; \theta)
    \Bigg]
    \Bigg|
    > \epsilon \sqrt{n}
    \right\}
    \end{aligned}
    \right] = 0 .
    \end{align*}

    \item (\textbf{Lindeberg-Feller condition}) 
        For every $\epsilon > 0$,
        \[
        \lim_{n \to \infty}
        \frac{1}{n}
        \sum_{i=1}^n
        E_{g_i}\!\left[
        \left\|
        \boldsymbol{\Omega}_n^{-1/2}\frac{\partial}{\partial \theta} 
        V_{\alpha,\beta,\gamma}(Y_i; \theta)
        \right\|^2
        \mathbb{I}\!\left\{
        \left\|
        \boldsymbol{\Omega}_n^{-1/2}\frac{\partial}{\partial \theta}
        V_{\alpha,\beta,\gamma}(Y_i; \theta)
        \right\|
        > \epsilon \sqrt{n}
        \right\}
        \right]
        = 0,
        \]
        where
        \[
        \boldsymbol{\Omega}_n 
        = \frac{1}{n}\sum_{i=1}^{n}\boldsymbol{K}_i ,
        \qquad 
        \boldsymbol{\Omega} 
        = \lim_{n\to\infty} \boldsymbol{\Omega}_n,
        \]
        and the limiting matrix $\boldsymbol{\Omega}$ is positive definite with
        \[
        \lambda_{\min}(\boldsymbol{\Omega}) > 0.
        \]
    \end{enumerate}

\medskip
\noindent
\textbf{Theorem 1.} \emph{
Under conditions (R1)–(R7), the following results hold:
\begin{enumerate}
    \item There exists a consistent sequence of roots $\hat{\theta}_n^{(\alpha,\beta,\gamma)}$ of the estimating equation \eqref{eq6} such that $\hat{\theta}_n^{(\alpha,\beta,\gamma)} \xrightarrow{p} \theta_g$.
    \item The Minimum Exponential-Polynomial Divergence estimator is asymptotically normal with
    \begin{equation}
    \sqrt{n}\,(\hat{\theta}_n^{(\alpha,\beta,\gamma)} - \theta_g)
    \xrightarrow{d} N_p(0, \boldsymbol{\Psi}^{-1}\boldsymbol{\Omega}\boldsymbol{\Psi}^{-1}).
    \label{eq14}
    \end{equation}
\end{enumerate}
}

\noindent
Thus, under standard smoothness and identifiability assumptions, the MEPDE for independent non-homogeneous observations is consistent and asymptotically normal, providing a unified robust inference framework within the exponential-polynomial divergence family.\\

\noindent\textbf{Remark.} \;
The proof of this theorem follows arguments analogous to those used in Theorem~3.1 of \citep{ghosh2013robust}, with appropriate modifications to incorporate the exponential–polynomial divergence structure.

\section{Exponential-Polynomial Divergence Information Criterion}\label{sec3}
\allowdisplaybreaks

Model selection criteria built upon divergence measures provide a systematic approach to weigh model fit against complexity, especially in situations where data contamination, heavy tails, or structural heterogeneity can distort likelihood-based criteria.  Divergence measures quantify how far a proposed model departs from the underlying data-generating distribution, and when the chosen divergence is itself resistant to outlying observations, the resulting information criterion naturally becomes more stable.  In this context, the exponential-polynomial divergence (EPD), introduced in Section~\ref{sec2}, serves as a versatile foundation for robust model selection.  Its three tuning parameters $(\alpha,\beta,\gamma)$ allow it to encompass several standard divergences—including the DPD, BED, and KL divergence—while offering additional flexibility to moderate the impact of outliers or model misspecification.

Building on this adaptability, we propose a robust information criterion based on the exponential-polynomial divergence, abbreviated as EPDIC.  This criterion extends the ideas underlying AIC and later divergence-based methods, including the DIC of \citep{mattheou2009model}, but incorporates the enhanced robustness properties of the EPD family.  Let
\[
\hat{\theta}^{(\alpha,\beta,\gamma)}
=
\arg\min_{\theta \in \Theta} \, H^{(\alpha,\beta,\gamma)}(\theta)
\]
represent the Minimum Exponential-Polynomial Divergence Estimator (MEPDE).  To construct an information criterion appropriate for EPD, one requires an asymptotically unbiased estimator of the expected overall divergence between the true density and the fitted parametric family, evaluated at $\hat{\theta}^{(\alpha,\beta,\gamma)}$.  Following the logic of AIC, this involves expanding the population divergence around the true parameter and deriving a model-complexity penalty whose structure depends on the curvature of the EPD functional.  The resulting penalty depends explicitly on the robustness parameters $(\alpha,\beta,\gamma)$, and it collapses to the usual AIC penalty when these parameters take the values corresponding to the Kullback-Leibler divergence.  Subsequently, we will describe the evolution of EPDIC.

In the present setting, the expected discrepancy between the true data-generating density $g$ and the fitted model $f_{\hat{\theta}}$ is quantified using the EPD functional
\[
\mathcal{D}_{\hat{\theta}}
\equiv 
H^{(\alpha,\beta,\gamma)}(g,f_{\hat{\theta}})
= \int_y D_{EP}(g(y),f_{\hat{\theta}}(y))\,dy.
\]

Let $\theta_0$ denote the true parameter such that $g=f_{\theta_0}$ under correct specification.  The criterion we wish to evaluate is the population quantity,
$E[\mathcal{D}(\hat{\theta})],$ which cannot be computed directly.  To relate it to observable components, we consider the empirical divergence functional
\[
\widehat{\mathcal{D}}(\theta)
= H_n^{(\alpha,\beta,\gamma)}(\theta),
\qquad 
\widehat{\mathcal{D}}(\hat\theta)
= H_n^{(\alpha,\beta,\gamma)}(\hat\theta),
\]
where $\hat{\theta}$ is the minimum exponential-polynomial divergence estimator (MEPDE).  We now employ a Taylor expansion of $\mathcal{D}(\theta)$ about $\theta_0$,
\begin{align}
\mathcal{D}(\hat{\theta})
&=
\mathcal{D}(\theta_{0})
+
(\hat{\theta}-\theta_{0})^{\!\top}
\left.
\frac{\partial}{\partial \theta}\,
\mathcal{D}(\theta)
\right|_{\theta=\theta_{0}}
\notag
\\[4pt]
&\quad+
\frac{1}{2}
(\hat{\theta}-\theta_{0})^{\!\top}
\left.
\frac{\partial^{2}}{\partial\theta\,\partial\theta^{\top}}
\mathcal{D}(\theta)
\right|_{\theta=\theta_{0}}
(\hat{\theta}-\theta_{0})
+
o\!\left(\|\hat{\theta}-\theta_{0}\|^{2}\right).
\label{eq15}
\end{align}
Since $\theta_0$ minimizes the population divergence,
\[
\left. \frac{\partial}{\partial{\theta}} \mathcal{D}(\theta) \right|_{\theta = \theta_0} = 0.
\]
We proceed further and get,
\begin{align}
E[ \mathcal{D}(\hat{\theta})]
&=
E[ \mathcal{D}(\theta_0) ]
+
\frac{1}{2}
E\!\left[(\hat{\theta}-\theta_{0})^{\!\top}
\left.
\frac{\partial^{2}}{\partial\theta\,\partial\theta^{\top}}
\mathcal{D}(\theta)
\right|_{\theta=\theta_{0}}
(\hat{\theta}-\theta_{0})\right]
+
o(n^{-1}).
\label{eq16}
\end{align}
Under the regularity conditions established for the EPD framework, the estimator $\hat{\theta}$ satisfies
\[
  \sqrt{n}\,(\hat{\theta} - \theta_0)
  \xrightarrow{d} N_p(0, \boldsymbol{\Psi}^{-1}\boldsymbol{\Omega}\boldsymbol{\Psi}^{-1}).
\]
Also note that, 
\[\left.
\frac{\partial^{2}}{\partial\theta\,\partial\theta^{\top}}
\mathcal{D}(\theta)
\right|_{\theta=\theta_{0}} = \boldsymbol{\Psi}.\]
Substituting this asymptotic behavior into the Taylor expansion gives 
\begin{align}
E[ \mathcal{D}(\hat{\theta}) ]
&=
\mathcal{D}(\theta_{0})
+
\frac{1}{2n}
tr\!\left(\boldsymbol{\Psi}E\!\left[(\hat{\theta}-\theta_{0})^{\!\top}
(\hat{\theta}-\theta_{0})\right]\right)
+
o(n^{-1})\notag \\[4pt]
&=
\mathcal{D}(\theta_{0})
+
\frac{1}{2n}
tr\!\left(\boldsymbol{\Omega} \boldsymbol{\Psi}^{-1}\right)
+
o(n^{-1}).
\label{eq17}
\end{align}
This result is directly analogous to the Takeuchi Information Criterion (TIC) expansion \citep{takeuchi1976distribution}.  Finally, replacing the unobservable $\mathcal{D}(\theta_{0})$ by its empirical counterpart $H_{n}^{(\alpha,\beta,\gamma)}(\hat{\theta})$ and multiplying both sides by $n$, yields the exponential-polynomial divergence information criterion
\begin{equation}
    \mathrm{EPDIC} \approx
    n\,H_{n}^{(\alpha,\beta,\gamma)}(\hat{\theta})
    + 
    tr\!\left(\boldsymbol{\Omega} \boldsymbol{\Psi}^{-1}\right).
    \label{eq18}
\end{equation}

\noindent The first term in \eqref{eq18} represents the empirical divergence and therefore measures the goodness of fit of the model under the chosen exponential-polynomial divergence.  The second term, 
$tr\!\left(\boldsymbol{\Omega} \boldsymbol{\Psi}^{-1}\right)$, serves as a model-complexity adjustment, playing the role of the asymptotic bias correction in the same spirit as Takeuchi’s information criterion.  Hence, EPDIC provides a flexible continuum of model-selection tools that unifies TIC and divergence measures from the exponential-polynomial family.

\section{Influence Function of the EPDIC}\label{sec4}
\allowdisplaybreaks

We now study the robustness properties of the proposed exponential-polynomial divergence information criterion (EPDIC) through its influence function, which quantifies the infinitesimal effect of a small contamination at a point on the value of the criterion. Throughout, we work under the independent non-homogeneous framework described in Section~\ref{sec2}.

It is worth emphasizing that the proposed EPDIC is defined as
\begin{equation*}
\mathrm{EPDIC}
=
n\,H_n^{(\alpha,\beta,\gamma)}(\hat{\theta})
+
\mathrm{tr}\!\left(\boldsymbol{\Omega}\boldsymbol{\Psi}^{-1}\right),
\end{equation*}
where $\hat{\theta}$ denotes the Minimum Exponential-Polynomial Divergence Estimator (MEPDE).

Let $G=(G_1,\ldots,G_n)$ denote the collection of true distributions, and consider a contaminated version of the $i^{th}$ component given by
\begin{equation}
G_{i,\varepsilon}
=
(1-\varepsilon)G_i + \varepsilon \Delta_y,
\label{eq19}
\end{equation}
where $\Delta_y$ is the degenerate distribution at $y$ and $\varepsilon \downarrow 0$.  The influence function of the EPDIC at the contamination point $y$ is defined as
\begin{equation}
\mathrm{IF}\!\left(y;\mathrm{EPDIC},G\right)
=
\left.
\frac{d}{d\varepsilon}
\mathrm{EPDIC}(G_1,\ldots,G_{i,\varepsilon},\ldots,G_n)
\right|_{\varepsilon=0}.
\label{eq20}
\end{equation}
The empirical divergence admits the functional representation
\begin{equation}
\mathcal{H}(\theta,G)
=
\frac{1}{n}\sum_{i=1}^n
\int
V_{\alpha,\beta,\gamma}(y;\theta)\,dG_i(y).
\label{eq21}
\end{equation}
Hence,
\begin{align}
\mathrm{IF}\!\left(y;nH_n(\hat{\theta}),G\right)
&=
n\Bigg[
V_{\alpha,\beta,\gamma}(y;\theta_g)
+
\left.
\frac{\partial H(\theta,G)}{\partial\theta}
\right|_{\theta=\theta_g}^{\!\top}
\mathrm{IF}(y;\hat{\theta},G)
\Bigg],
\label{eq22}
\end{align}
where $\theta_g$ denotes the population minimizer of the divergence. Since $\theta_g$ satisfies
\begin{equation}
\left.
\frac{\partial H(\theta,G)}{\partial\theta}
\right|_{\theta=\theta_g}
=0,
\label{eq23}
\end{equation}
we obtain
\begin{equation}
\mathrm{IF}\!\left(y;nH_n(\hat{\theta}),G\right)
=
n\,V_{\alpha,\beta,\gamma}(y;\theta_g).
\label{eq24}
\end{equation}
The penalty component $\mathrm{tr}\!\left(\boldsymbol{\Omega}\boldsymbol{\Psi}^{-1}\right)$ depends on the underlying distribution only through second-order moments.  Under standard smoothness and moment conditions, its influence function is of order $O(1)$ and is therefore negligible compared to the $O(n)$ divergence contribution.  Thus,
\begin{equation}
\mathrm{IF}\!\left(
y;
\mathrm{tr}(\boldsymbol{\Omega}\boldsymbol{\Psi}^{-1}),
G
\right)
=
O(1).
\label{eq25}
\end{equation}
Combining the above results, the influence function of the classical EPDIC is given by
\begin{equation}
\mathrm{IF}\!\left(y;\mathrm{EPDIC},G\right)
=
n\,V_{\alpha,\beta,\gamma}(y;\theta_g)
+
O(1).
\label{eq26}
\end{equation}

\noindent Since $V_{\alpha,\beta,\gamma}(y;\theta)$ is bounded for $\alpha>0$ and $\gamma>0$, the influence function of the classical EPDIC is bounded, confirming its robustness.  In the limiting case $(\alpha,\beta,\gamma)\to(0,0,0)$, the divergence reduces to the negative log-likelihood, yielding an unbounded influence function and recovering the non-robust behavior of likelihood-based criteria such as AIC and TIC.

The robustness properties of information criteria can be assessed through their influence functions (IFs), which quantify the effect of an infinitesimal contamination on the criterion value.  Conventional information criteria like AIC are directly constructed from the log-likelihood evaluated at the maximum likelihood estimator. Consequently, their influence functions inherit the unbounded nature of the score and log-likelihood contributions, rendering AIC highly sensitive to outlying observations.  In particular, a single gross error can exert an arbitrarily large impact on these criteria, leading to potentially unstable model selection decisions.  In contrast, the proposed EPDIC replaces the log-likelihood component with a robust divergence-based objective, leading to bounded score-type contributions.  This modification yields a bounded influence function for EPDIC, ensuring that the effect of anomalous observations is controlled.  Consequently, EPDIC exhibits reduced gross-error sensitivity and improved local robustness compared to AIC and DIC.  This boundedness property makes EPDIC particularly well-suited for model selection in the presence of mild deviations from the assumed model or potential data contamination.

\section{Optimal Tuning Parameter}\label{sec5}
\allowdisplaybreaks

The performance of the Minimum Exponential-Polynomial Divergence Estimator (MEPDE) critically depends on the appropriate choice of the robustness parameters $(\alpha,\beta,\gamma)$.  While these parameters regulate the trade-off between efficiency and robustness, their optimal selection is non-trivial.  Classical approaches such as cross-validation or asymptotic mean squared error minimization require either repeated model fitting or explicit evaluation of asymptotic covariance matrices, both of which may be computationally demanding within the exponential-polynomial framework.

To address this issue, we adopt the generalized score matching principle introduced by \citep{hyvarinen2005estimation}, which provides an alternative estimation and model selection mechanism that does not rely on the normalizing constant of the model density.  In particular, we exploit the formulation of generalized score matching in Euclidean space for independent observations, as detailed in \citep{xu2025generalized}.  This approach offers a computationally attractive and theoretically justified criterion for selecting the optimal tuning parameters of the exponential-polynomial divergence estimator.

Let $Y_1,\ldots,Y_n \in \mathbb{R}^d$ be independent observations with true densities $g_i$, modeled by parametric densities $f_i(\cdot;\theta)$.  Denote by 
\[
p_i(y;\theta) = f_i(y;\theta)
\]
the model density.  The generalized score matching criterion is based on the Fisher divergence between the true density and the model density, defined as
\[
D_{\mathrm{SM}}(g,p)
=
\frac{1}{n}\sum_{i=1}^n 
E_{g_i}
\Big[
\Big\|
\nabla_i \log g_i(Y_i)
-
\nabla_i \log p_i(Y_i;\theta)
\Big\|^2
\Big],
\]
where $\nabla_i$ denotes differentiation with respect to $Y_i$.

Under mild regularity conditions, this divergence decomposes into a term independent of $\theta$ and a model-dependent component.  Consequently, minimizing the Fisher divergence reduces to minimizing the empirical generalized score matching objective
\begin{equation}
d_{\mathrm{SM}}(\theta)
=
\frac{1}{n}\sum_{i=1}^n
\rho_{\mathrm{SM},i}(Y_i;\theta),
\label{eq27}
\end{equation}
where
\begin{equation}
\rho_{\mathrm{SM},i}(Y_i;\theta)
=
2\sum_{j=1}^d
\frac{\partial^2}{\partial y_{ij}^2}
\log f_i(Y_i;\theta)
+
\sum_{j=1}^d
\left(
\frac{\partial}{\partial y_{ij}}
\log f_i(Y_i;\theta)
\right)^2.
\label{eq28}
\end{equation}
An important advantage of this formulation is that it depends only on derivatives of the log-density with respect to the data, thereby eliminating the need to evaluate any normalizing constant.  This feature is particularly beneficial in divergence-based estimation frameworks.  For each fixed triplet $(\alpha,\beta,\gamma)$, let
\[
\hat{\theta}^{(\alpha,\beta,\gamma)}
=
\arg\min_{\theta}
H_n^{(\alpha,\beta,\gamma)}(\theta)
\]
denote the corresponding MEPDE.  Substituting this estimator into the generalized score matching objective \eqref{eq27}, we define the tuning-selection criterion
\begin{equation}
\mathcal{S}_n(\alpha,\beta,\gamma)
=
d_{\mathrm{SM}}\!\left(
\hat{\theta}^{(\alpha,\beta,\gamma)}
\right).
\label{eq29}
\end{equation}
The optimal tuning parameters are then obtained as
\begin{equation}
(\hat{\alpha},\hat{\beta},\hat{\gamma})
=
\arg\min_{\alpha,\beta,\gamma}
\mathcal{S}_n(\alpha,\beta,\gamma).
\label{eq30}
\end{equation}
This procedure selects the divergence parameters that yield the smallest Fisher divergence between the fitted model and the underlying data-generating mechanism.  Unlike approaches based on asymptotic MSE minimization, this method does not require explicit computation of the asymptotic covariance matrices $\boldsymbol{\Psi}$ and $\boldsymbol{\Omega}$.  Moreover, it avoids repeated pilot updates, as required in iterative Warwick and Jones-type procedures \citep{basak2021optimal}.  Once the optimal tuning parameters $(\hat{\alpha},\hat{\beta},\hat{\gamma})$ are determined via \eqref{eq30}, the corresponding MEPDE 
\[
\hat{\theta}^{*}
=
\hat{\theta}^{(\hat{\alpha},\hat{\beta},\hat{\gamma})}
\]
is used to evaluate the EPDIC, defined in Section~\ref{sec3}.

\section{Numerical Experiments}\label{sec6}
\allowdisplaybreaks

This section presents an extensive set of numerical experiments designed to examine the finite-sample performance and practical relevance of the proposed methodologies.  Comprehensive simulation studies are conducted to assess the behavior of the estimators and associated information criteria under a variety of data-generating scenarios.  In addition, two real-world applications are analyzed to illustrate the practical applicability of the proposed framework. The first application concerns a linear mixed-effects panel data model, while the second focuses on a neural network setting, thereby demonstrating the flexibility and effectiveness of the proposed approach across both classical statistical models and modern machine learning frameworks.

\subsection{\textbf{Simulation Study}}
\allowdisplaybreaks

An extensive Monte Carlo simulation study is conducted to investigate the robustness of the proposed exponential-polynomial divergence information criteria.  All results reported in this subsection are based on $1000$ independent Monte Carlo replications.  We consider a multiple linear regression model with $p=5$ covariates, given by
\begin{equation}
y_i = \boldsymbol{x}_i^{\top}\boldsymbol{\beta} + \varepsilon_i, \quad i=1,\ldots,n,
\label{eq31}
\end{equation}
where $\boldsymbol{x}_i = (x_{i1},\ldots,x_{i5})^{\top}$ denotes the vector of explanatory variables, $\boldsymbol{\beta}$ is the vector of regression coefficients, and $\varepsilon_i$ represents the random error term.  The response variable $y_i$ is generated from a Gaussian distribution, ensuring compatibility with the likelihood-based and divergence-based estimation frameworks.  While, the covariate vectors $\boldsymbol{x}_i$ are generated from a multivariate normal distribution with mean vector $\boldsymbol{0}$ and a covariance matrix $\boldsymbol{\Sigma}$ having an autoregressive structure, that is,
\[
\Sigma_{jk} = \rho^{|j-k|}, \quad j,k = 1,\ldots,5,
\]
with $\rho = 0.5$, allowing for moderate correlation among the covariates.  The error terms $\varepsilon_i$ are independently generated from a normal distribution with mean zero and variance $\sigma^2 = 1$.  The true regression coefficient vector is fixed as
\[
\boldsymbol{\beta}_0 = (1.5, -1.0, 0.8, 0.5, -0.7)^{\top}.
\]

Subsequently, the simulation procedure is implemented under a single sample size, namely $n = 150$.  To examine the robustness properties of the estimators, two distinct contamination schemes are considered.  In the first scheme, contamination is introduced in the error (disturbance) terms $\varepsilon_i$, where a proportion $\delta \in (0.052, 0.093, 0.134)$ of the errors is replaced by values generated from a $\text{Normal}(10.6, 1)$ distribution.  This mechanism induces vertical outliers in the response variable while leaving the design matrix unaffected.  In the second scheme, contamination is introduced in the explanatory variables $\boldsymbol{x}_i$.  Specifically, a proportion $\delta \in (0.058, 0.099, 0.140)$ of the covariate observations is replaced by values drawn from a $\text{Shifted-Normal}(45.6, 6.3)$ distribution.  This setup generates leverage-type outliers in the design matrix.  Taken together, these contamination mechanisms introduce both moderate and severe deviations from the assumed model structure.

Next, parameter estimation is performed using three competing approaches: the classical maximum likelihood estimator (MLE), the density power divergence (DPD) estimator, and the exponential-polynomial divergence (EPD) estimator.  These estimators are obtained by minimizing their respective objective functions through a sequential convex programming (SCP) algorithm (see \citep{dinh2010local}).  Within this framework, each nonlinear objective function is locally approximated by a convex surrogate and solved iteratively.  At every iteration, a convex subproblem is constructed using first-order derivative information together with an appropriate step-size control to ensure numerical stability and monotonic descent.  The parameter vector is updated according to the solution of the convex approximation, and the process continues until convergence.  The algorithm is initialized at the ordinary least squares (OLS) estimates, and convergence is declared when the relative change in both the parameter estimates and the objective function value falls below a pre-specified tolerance threshold.

Furthermore, the optimal tuning parameters for both the DPD and EPD estimators are determined using the generalized score matching procedure under each contamination scheme.  This data-driven strategy identifies the tuning parameters that achieve an appropriate balance between efficiency and robustness by minimizing an empirical criterion derived from the corresponding score equations.  The resulting optimal tuning parameter values under the different contamination scenarios are summarized in \textcolor{blue}{\hyperref[tab1]{Table~\ref{tab1}}}.

\begin{table}[H]
\centering
\fontsize{9}{11}\selectfont
\caption{Optimal tuning parameters for DPD and EPD under different contamination schemes}
\label{tab1}
\renewcommand{\arraystretch}{1.2}
\begin{tabular}{cccccc}
\hline
 & $\boldsymbol{\delta}$ 
& $\boldsymbol{\gamma}$ (DPD) 
& $\boldsymbol{\alpha}$ (EPD) 
& $\boldsymbol{\beta}$ (EPD) 
& $\boldsymbol{\gamma}$ (EPD) \\
\hline
\textbf{Pure} & 0.000 
& $0.95$ 
& $0.10$ 
& $0.70$ 
& $0.30$ \\
\hline
& $0.052$ 
& $0.25$ 
& $0.10$ 
& $0.60$ 
& $0.70$ \\

\textbf{Cont. 1}  & $0.093$ 
& $0.35$ 
& $0.10$ 
& $0.70$ 
& $0.30$ \\

 & $0.134$ 
& $0.40$ 
& $0.40$ 
& $0.70$ 
& $0.60$ \\
\hline
& $0.058$ 
& $0.15$ 
& $0.10$ 
& $0.70$ 
& $0.90$ \\

\textbf{Cont. 2}  & $0.099$ 
& $0.25$ 
& $0.10$ 
& $0.60$ 
& $0.70$ \\

 & $0.140$ 
& $0.20$ 
& $0.10$ 
& $0.70$ 
& $0.90$ \\
\hline
\end{tabular}
\end{table}

\noindent Employing these optimal tuning parameters, the parameter estimates are ultimately derived using the MLE, DPDE, and EPDE methods, noting that MLE does not require any tuning parameter.  The corresponding numerical results, evaluated across various contamination scenarios, are reported in \textcolor{blue}{\hyperref[tab2]{Table~\ref{tab2}}}.

\begin{table}[htb!]
\centering
\fontsize{9}{11}\selectfont
\caption{Estimated values of the parameters under pure and contamination schemes.}
\label{tab2}
\renewcommand{\arraystretch}{1.2}
\vspace{0.2cm}
\begin{tabular}{lccccc}
\hline
\textbf{Scheme} & $\boldsymbol{\delta}$ & \textbf{Parameter} & \textbf{MLE} & \textbf{DPDE} & \textbf{EPDE} \\
\hline

\multirow{5}{*}{\textbf{Pure}} 
& \multirow{5}{*}{$\boldsymbol{0.000}$}
& $\beta_1$ & 1.495108 & 1.494928 & 1.494640 \\
& & $\beta_2$ & -0.995442 & -0.995126 & -0.994971 \\
& & $\beta_3$ & 0.799332 & 0.799579 & 0.799965 \\
& & $\beta_4$ & 0.502114 &  0.502597 & 0.503010 \\
& & $\beta_5$ & -0.701112 & -0.700919 & -0.700425 \\
\hline

\multirow{15}{*}{\textbf{Cont. 1}} 
& \multirow{5}{*}{$\boldsymbol{0.052}$}
& $\beta_1$ & 1.569221 & 1.520784 & 1.514209 \\
& & $\beta_2$ & -1.018774 & -0.981226 & -0.989350 \\
& & $\beta_3$ & 0.761441 & 0.772438 & 0.785769 \\
& & $\beta_4$ & 0.548332 & 0.520714 & 0.513276 \\
& & $\beta_5$ & -0.731114 & -0.706512 & -0.701868 \\
\cline{2-6}

& \multirow{5}{*}{$\boldsymbol{0.093}$}
& $\beta_1$ & 1.628114 & 1.526884 & 1.517182 \\
& & $\beta_2$ & -1.061884 & -0.978112 & -0.980029 \\
& & $\beta_3$ & 0.726441 & 0.759331 & 0.762461 \\
& & $\beta_4$ & 0.589114 & 0.530774 & 0.521705 \\
& & $\beta_5$ & -0.768552 & -0.710221 & -0.703512 \\
\cline{2-6}

& \multirow{5}{*}{$\boldsymbol{0.134}$}
& $\beta_1$ & 1.702441 & 1.539882 & 1.525483 \\
& & $\beta_2$ & -1.112774 & -0.964221 & -0.973818 \\
& & $\beta_3$ & 0.691118 & 0.741115 & 0.757822 \\
& & $\beta_4$ & 0.629441 & 0.542997 & 0.532960 \\
& & $\beta_5$ & -0.804118 & -0.715438 & -0.706917 \\
\hline

\multirow{15}{*}{\textbf{Cont. 2}} 
& \multirow{5}{*}{$\boldsymbol{0.058}$}
& $\beta_1$ & 1.528114 & 1.502398 & 1.501286 \\
& & $\beta_2$ & -1.006218 & -0.998402 & -0.999017 \\
& & $\beta_3$ & 0.819552 & 0.801669 & 0.800941 \\
& & $\beta_4$ & 0.515008 & 0.508681 & 0.501237 \\
& & $\beta_5$ & -0.721933 & -0.703284 & -0.700739 \\
\cline{2-6}

& \multirow{5}{*}{$\boldsymbol{0.099}$}
& $\beta_1$ & 1.566227 & 1.511903 & 1.507549 \\
& & $\beta_2$ & -1.023915 & -0.979114 & -0.989718 \\
& & $\beta_3$ & 0.845661 & 0.810492 & 0.808795 \\
& & $\beta_4$ & 0.538104 & 0.512961 & 0.509558 \\
& & $\beta_5$ & -0.748991 & -0.710728 & -0.702381 \\
\cline{2-6}

& \multirow{5}{*}{$\boldsymbol{0.140}$}
& $\beta_1$ & 1.621332 & 1.517884 & 1.511344 \\
& & $\beta_2$ & -1.056781 & -0.968331 & -0.975795 \\
& & $\beta_3$ & 0.892114 & 0.819612 & 0.813201 \\
& & $\beta_4$ & 0.579442 & 0.526893 & 0.513679 \\
& & $\beta_5$ & -0.789661 & -0.718947 & -0.710308 \\
\hline

\end{tabular}
\end{table}

Based on the estimated parameter values from each competing method, the corresponding information criteria—namely, the maximum likelihood information criterion (MLIC), the density power divergence information criterion (DPDIC), and the proposed exponential-polynomial divergence information criterion (EPDIC)---are then evaluated under both pure and contaminated data-generating mechanisms.  In this unified framework, the criteria are computed by combining the empirical likelihood-, DPD-, and EPD-based objective functions evaluated at their respective estimators with the associated asymptotic penalty terms that account for model complexity.  This formulation enables a coherent comparison of likelihood-based and divergence-based model selection strategies.

\begin{figure}[!t]
\centering
\includegraphics[width=0.63\linewidth]{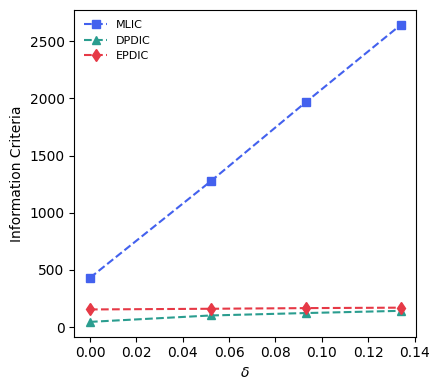}
\caption{Information criteria under Contamination scheme 1.}
\label{fig1}
\end{figure}

\begin{figure}[!t]
\centering
\includegraphics[width=0.63\linewidth]{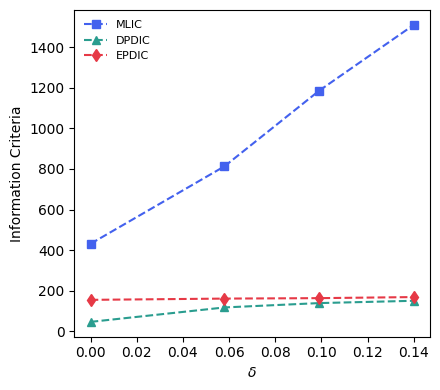}
\caption{Information criteria under Contamination scheme 2.}
\label{fig2}
\end{figure}

\textcolor{blue}{\hyperref[fig1]{Figure~\ref{fig1}}} illustrates the information criterion values for each estimation method under Scheme~1 across varying contamination proportions $\delta$, while \textcolor{blue}{\hyperref[fig2]{Figure~\ref{fig2}}} displays the corresponding results under Scheme~2.  The comparative behavior of MLIC, DPDIC, and EPDIC as $\delta$ increases in each scheme provides a clear assessment of their relative stability and robustness in the presence of model deviations, thereby highlighting the effect of different contamination structures on model selection performance.

Consistent with the patterns displayed in \textcolor{blue}{\hyperref[fig1]{Figure~\ref{fig1}}} and \textcolor{blue}{\hyperref[fig2]{Figure~\ref{fig2}}}, the rate of increase in the information criterion values as the contamination proportion $\delta$ rises clearly distinguishes the three approaches.  Under pure data, all criteria remain stable; however, once contamination is introduced, their trajectories diverge markedly. The MLIC exhibits the steepest escalation with increasing $\delta$ in both Scheme~1 (error contamination) and Scheme~2 (leverage contamination), indicating pronounced sensitivity to model deviations.  The rapid growth of MLIC reflects the well-known vulnerability of likelihood-based criteria to both vertical and leverage-type outliers.

In contrast, DPDIC demonstrates a more moderate and controlled increase as contamination intensifies.  Although its values do rise with $\delta$, the progression is substantially smoother than that of MLIC, suggesting improved resistance to departures from model assumptions.  Among the three, EPDIC shows the smallest incremental change across contamination levels in both schemes.  Its comparatively minimal rise as $\delta$ increases highlights a high degree of stability, thereby underscoring the robustness of the exponential-polynomial divergence framework.  Overall, the relative magnitudes of increase—largest for MLIC, moderate for DPDIC, and smallest for EPDIC—provide coherent empirical evidence of the superior robustness of the divergence-based information criteria, particularly EPDIC, under different contaminated settings.

\subsection{\textbf{Real data Analysis}}
\allowdisplaybreaks

\subsubsection{Application: Linear Mixed-Effects Panel data Models}\label{5}

We consider a linear mixed-effects panel data model of the form
\begin{equation}
    y_{it} = x_{it}^{\top}\beta + z_{it}^{\top}\alpha_i + u_{it},
    \qquad
    i=1,\ldots,n,\; t=1,\ldots,m,
\label{eq32}
\end{equation}
with $\alpha_i \sim \mathcal{N}_p(0,\Sigma_\alpha)$ and $u_{it}\sim \mathcal{N}(0,\sigma_u^2)$ independently.  The observations are assumed to be independent but non-homogeneous across individuals, allowing the covariance structure to vary with $i$ through the individual-specific design matrices.  By stacking the observations over time for each individual, the model can be expressed as
\[
\boldsymbol{y}_i = \boldsymbol{X}_i\beta + \boldsymbol{\varepsilon}_i,
\qquad
\boldsymbol{\varepsilon}_i \sim \mathcal{N}_m(0,\boldsymbol{\Omega}_i),
\]
where
\[
\boldsymbol{\Omega}_i = \boldsymbol{Z}_i\Sigma_\alpha\boldsymbol{Z}_i^\top + \sigma_u^2\boldsymbol{I}_m .
\]
Hence, the marginal density of $\boldsymbol{y}_i$ is
\begin{equation}
    f_i(\boldsymbol{y}_i;\boldsymbol{\theta})
    =
    (2\pi)^{-m/2}|\boldsymbol{\Omega}_i|^{-1/2}
    \exp\!\left\{
    -\frac{1}{2}
    (\boldsymbol{y}_i-\boldsymbol{X}_i\beta)^\top
    \boldsymbol{\Omega}_i^{-1}
    (\boldsymbol{y}_i-\boldsymbol{X}_i\beta)
    \right\},
\label{eq33}
\end{equation}
where $\boldsymbol{\theta}=(\beta^\top,\mathrm{vec}(\Sigma_\alpha)^\top,\sigma_u^2)^\top$.

With this general formulation in place, we now illustrate the application of the linear mixed-effects panel model using a real-world dataset that naturally aligns with its structure.  In particular, we consider the \textbf{Panel Data of Individual Wages} dataset, a well-known benchmark dataset in panel data econometrics.  This dataset is drawn from the \textbf{Panel Study of Income Dynamics (PSID)} and contains longitudinal wage information for $59$5 individuals observed over the period $1976-1982$ in the United States, resulting in $4165$ total observations arranged as a balanced panel.

The primary response variable in this study is the logarithm of wage (\textit{lwage}), which is commonly used in labor economics analyses to stabilize variance and interpret coefficients in percentage terms. The set of explanatory variables includes both continuous and categorical covariates reflecting individual characteristics and employment conditions, such as years of education (\textit{ed}), years of full-time work experience (\textit{exp}), weeks worked (\textit{wks}), union membership (\textit{union}), marital status (\textit{married}), industry indicators (\textit{bluecol, ind}), regional indicators (\textit{south, smsa}), and demographic attributes including sex and race (\textit{black}).  These variables are incorporated in the model as fixed effects, representing systematic influences on wage levels that are assumed to be common across individuals.

To account for unobserved heterogeneity that cannot be explained by the observed covariates—such as innate ability, motivation, or persistent individual-specific productivity—we introduce individual-level random effects.  These random effects capture subject-specific deviations from the overall wage-covariate relationship and are assumed to follow a normal distribution with mean zero and unknown variance.  The inclusion of random effects is particularly appropriate here because the dataset contains repeated observations for each individual over multiple years, enabling the model to separate within-individual temporal variation from between-individual variability.

The wage panel dataset is well-suited to the proposed linear mixed-effects framework, as it provides a balanced longitudinal structure with rich socio-economic covariates and repeated observations for each individual, enabling reliable estimation of both fixed and random effects.  To assess multivariate extremeness in the joint distribution of the covariates, we employ the Minimum Covariance Determinant (MCD) estimator.  It is a widely used, robust multivariate outlier detection method that seeks the subset of observations with the smallest determinant of the covariance matrix, thereby providing high-breakdown estimates of location and scatter that are not unduly influenced by outliers.  The robust Mahalanobis distances derived from the MCD estimate are used to detect multivariate outliers, as recommended in the robust statistics literature (e.g., \citep{rousseeuw1999fast}, \citep{hubert2003pca}).  The MCD-based diagnostic confirms that, despite some individually extreme values, there are no observations that simultaneously deviate in the multivariate covariate space to an extent requiring exclusion. Therefore, no observations have been removed, and robust estimation procedures are used to mitigate potential influence from atypical observations.  In addition, the dataset is verified to contain no missing values. 

We employ MLE, DPDE, and EPDE for estimating the model parameters.  From the perspective of exponential-polynomial divergence, the parameter estimation problem reduces to minimizing an empirical divergence criterion.  Specifically, we consider the objective function
\begin{equation}
H_n^{(\alpha,\beta,\gamma)}(\boldsymbol{\theta})
=
\frac{1}{n}\sum_{i=1}^{n}
V_{\alpha,\beta,\gamma}(\boldsymbol{y}_i;\boldsymbol{\theta}),
\label{eq34}
\end{equation}
where the contribution from the $i$th subject is captured by
\begin{align}
V_{\alpha,\beta,\gamma}(\boldsymbol{y}_i;\boldsymbol{\theta})
&=
\frac{\beta}{\alpha^2}
\int_{\mathbb{R}^m}
\left[
e^{\alpha f_i(\boldsymbol{y};\boldsymbol{\theta})}
\big(\alpha f_i(\boldsymbol{y};\boldsymbol{\theta}) - 1\big) + 1
\right] d\boldsymbol{y}
\nonumber \\
&\quad
+ \frac{1-\beta}{\gamma}
\int_{\mathbb{R}^m}
\left[
f_i^{\,\gamma+1}(\boldsymbol{y};\boldsymbol{\theta}) - f_i(\boldsymbol{y};\boldsymbol{\theta})
\right] d\boldsymbol{y}
\nonumber \\
&\quad
-
\left[
\frac{\beta}{\alpha}
\left(e^{\alpha f_i(\boldsymbol{y}_i;\boldsymbol{\theta})} - 1\right)
+
\frac{1-\beta}{\gamma}
\left((\gamma+1)f_i(\boldsymbol{y}_i;\boldsymbol{\theta}) - 1\right)
\right].
\label{eq35}
\end{align}
To obtain stable starting values for the iterative estimation procedures, the fixed-effect parameters are first initialized using the ordinary least squares (OLS) estimator applied to the pooled regression model.  These preliminary estimates serve as initial values for the subsequent divergence-based optimization procedures.  Based on these initial estimates, the optimal tuning parameters for the divergence-based estimators are determined using the generalized score matching method.  For DPDE, the optimal tuning parameter is obtained as \[\gamma = 0.50, \]
while for EPDE, the optimal set of tuning parameters is found to be \[(\alpha, \beta, \gamma) = (0.1, 0.3, 0.3).\]
Using these optimal values, the fixed-effects model parameters are estimated.

Subsequently, to identify the most relevant explanatory variables and reduce the computational burden associated with evaluating all possible regression specifications, we incorporate a LASSO-type penalty into the likelihood-, DPD-, and EPD-based objective functions.  This penalized estimation step performs variable selection by shrinking the coefficients of less informative covariates toward zero.  The procedure is particularly useful in the present context because calculating the information criteria for the full model with all available covariates would require evaluating an excessively large number of candidate models.  The penalized estimation results reveal a consistent subset of covariates that remain relevant across all three objective functions.  The variables selected by the LASSO procedure are
\[(\textit{bluecol}, \textit{smsa}, \textit{married}, \textit{sex}, \textit{union}, \textit{black}).\]
Accordingly, the model-selection analysis is restricted to these six covariates.  By examining every non-empty subset of these variables, we obtain $(2^6 - 1 = 63)$ candidate regression models.  For each of these $63$ candidate models, the corresponding information criteria---MLIC, DPDIC, and EPDIC---are computed.  The models are then ranked in ascending order according to the magnitude of each criterion.  From the three ranked lists obtained under MLIC, DPDIC, and EPDIC, the top fifteen candidate models are extracted.  These lists are thereupon consolidated to form a combined set of candidate models, within which several models appear repeatedly across the three criteria.

To identify the model specifications that receive the most consistent support across the three criteria, we compute the frequency with which each candidate model appears among the top-ranked models.  Based on this frequency, a final set of five leading candidate models is selected.  \textcolor{blue}{\hyperref[tab3]{Table~\ref{tab3}}} presents these models along with their selection frequencies and the corresponding values of EPDIC, DPDIC, and MLIC.

\begin{table}[H]
\centering
\fontsize{9}{11}\selectfont
\caption{Top candidate models based on consolidated information-criterion rankings.}
\label{tab3}
\vspace{-5pt}
\begin{tabular}{p{5cm} c c r r r}
\hline
\textbf{Model} & \textbf{Freq.} & \textbf{Sel. Freq.} & \textbf{EPDIC} & \textbf{DPDIC} & \textbf{MLIC} \\
\hline
bluecol, sex, black & 3 & 1.000 & 1435.263 & 229.14054 & 11666.50 \\
smsa, married, sex & 3 & 1.000 & 1892.022 & 370.21159 & 11101.19 \\
bluecol, smsa, sex, union & 2 & 0.667 & 1669.711 & 246.01321 & 21262.38 \\
bluecol, smsa, sex, black & 2 & 0.667 & 1806.677 & 201.14246 & 21313.76 \\
bluecol, smsa, sex, union, black & 2 & 0.667 & 1716.873 & 253.14976 & 21249.34 \\
\hline
\end{tabular}
\end{table}

The results indicate that several models receive consistent support across multiple information criteria, suggesting that the identified covariates play a significant role in explaining wage variation.  In particular, models involving occupation type (\textit{bluecol}), regional indicator (\textit{smsa}), marital status, gender, union membership, and race repeatedly appear among the leading specifications.  This consolidated ranking approach provides a robust mechanism for identifying economically meaningful wage determinants while avoiding overfitting and excessive model complexity.

\subsubsection{Application: Neural Network Models}\label{6}

Beyond classical parametric and panel-data models, the proposed divergence-based estimation and model-selection framework can be naturally extended to modern machine-learning architectures.  In this subsection, we illustrate its applicability through supervised classification using feed-forward neural networks (FFNNs), which are widely employed nonlinear predictive models in statistical learning.

Let $\{(\boldsymbol{x}_i, y_i)\}_{i=1}^{N}$ denote a collection of independent observations, where $\boldsymbol{x}_i \in \mathbb{R}^{d}$ represents a $d$-dimensional feature vector and $y_i \in \{0,1\}$ denotes a binary class label indicating the operational state of a machine.  In particular,
\[
y_i =
\begin{cases}
1, & \text{if a machine failure occurs},\\
0, & \text{otherwise}.
\end{cases}
\]

To illustrate the proposed framework, we utilize the \textbf{AI4I 2020 Predictive Maintenance Dataset}, a widely used benchmark dataset in predictive maintenance and machine-learning studies. The dataset contains information on $10,000$ industrial machine instances along with several operational and environmental variables that influence machine performance and reliability.

The primary objective is to predict whether a machine is likely to fail based on a set of sensor-based covariates.  The explanatory variables considered in the analysis include air temperature, process temperature, rotational speed, torque, and tool wear, together with a categorical machine-type indicator reflecting the product quality level.  These variables represent key operational characteristics that influence mechanical stress, thermal conditions, and tool degradation in manufacturing environments.

Prior to model estimation, all continuous covariates are standardized to ensure numerical stability and comparable scaling across features.  The categorical machine-type variable is encoded using dummy variables.  The dataset is verified to contain no missing values.

To examine potential multivariate extremeness among the covariates, we employ the Minimum Covariance Determinant (MCD) estimator.  This robust method identifies subsets of observations with minimal covariance determinant and provides high-breakdown estimates of multivariate location and scatter.  The corresponding robust Mahalanobis distances are used to detect multivariate outliers.  The diagnostic analysis indicates that although some observations exhibit individually extreme sensor values, none deviate substantially in the joint covariate space.  Consequently, no observations are removed, and robust estimation procedures are retained to mitigate potential influence from outlying measurements.

Instead of selecting subsets of covariates, model selection in neural networks involves choosing among alternative network architectures.  Accordingly, we fix a finite collection of candidate architectures,
\[
\mathcal{A}=\{A_1,A_2,\ldots,A_m\},
\]
each differing in structural complexity but using all available covariates as inputs.  The architectural variation is constructed by altering the number of hidden layers and the number of neurons per layer.  Specifically, we consider
\begin{itemize}
\item Number of hidden layers $L \in \{1,2\}$,
\item Number of hidden neurons per layer $H \in \{2,3\}$.
\end{itemize}

\noindent This specification yields a finite set of candidate architectures corresponding to all possible $(L,H)$ combinations.  Each architecture employs the same standardized input feature space and a softmax output layer for binary classification.  Thus, the candidate models differ only in internal network complexity while retaining identical covariate information.  For a given architecture $A_j$, the predicted class probability for observation $i$ is expressed as
\[
\hat{p}_i =
\mathrm{Softmax}\!\left(
f_{A_j}(\boldsymbol{x}_i;\boldsymbol{\theta}_j)
\right),
\]
where $f_{A_j}(\cdot)$ denotes the feed-forward transformation determined by architecture $A_j$, and $\boldsymbol{\theta}_j$ represents the collection of weights and bias parameters associated with the network.  Thus, $\hat{p}_i \in (0,1)$ represents the predicted probability of machine failure $(y_i=1)$.  The corresponding Bernoulli model implied by the network is given by
\[
f_i(y_i;\boldsymbol{\theta}_j)
= \hat{p}_i^{\,y_i} (1-\hat{p}_i)^{\,1-y_i}, 
\quad y_i \in \{0,1\}.
\]

Using the EPD framework, the loss function for neural network estimation is defined as
\begin{equation}
\mathcal{L}^{(\alpha,\beta,\gamma)}(\boldsymbol{\theta}_j)
=
\frac{1}{N}\sum_{i=1}^{N}
V_{\alpha,\beta,\gamma}(y_i;\boldsymbol{\theta}_j),
\label{eq36}
\end{equation}
where the samplewise contribution $V_{\alpha,\beta,\gamma}(\cdot)$ is given by
\begin{align}
V_{\alpha,\beta,\gamma}(y_i;\boldsymbol{\theta}_j)
&=
\frac{\beta}{\alpha^2}
\sum_{y \in \{0,1\}}
\left[
e^{\alpha f_i(y;\boldsymbol{\theta}_j)}
\big(\alpha f_i(y;\boldsymbol{\theta}_j) - 1\big) + 1
\right]
\nonumber \\
&\quad
+ \frac{1-\beta}{\gamma}
\sum_{y \in \{0,1\}}
\left[
f_i^{\gamma+1}(y;\boldsymbol{\theta}_j) - f_i(y;\boldsymbol{\theta}_j)
\right]
\nonumber \\
&\quad
-
\left[
\frac{\beta}{\alpha}
\left(e^{\alpha f_i(y_i;\boldsymbol{\theta}_j)} - 1\right)
+
\frac{1-\beta}{\gamma}
\left((\gamma+1)f_i(y_i;\boldsymbol{\theta}_j) - 1\right)
\right].
\label{eq37}
\end{align}

\noindent
Since the response is binary, the summations are taken over $y \in \{0,1\}$, with
\[
f_i(1;\boldsymbol{\theta}_j) = \hat{p}_i,
\quad
f_i(0;\boldsymbol{\theta}_j) = 1 - \hat{p}_i.
\]
The neural network parameters are then estimated by minimizing the above loss function:
\begin{equation}
\hat{\boldsymbol{\theta}}_j
=
\arg\min_{\boldsymbol{\theta}_j}
\mathcal{L}{(\alpha,\beta,\gamma)}(\boldsymbol{\theta}_j).
\label{eq38}
\end{equation}
Similarly, the corresponding estimators are obtained by minimizing the respective loss functions over the parameter space.  

To obtain stable starting values for the optimization procedure, the network weights are initialized using standard random initialization methods commonly employed in neural network training.  Based on these initial values, the optimal tuning parameters for the divergence-based estimators are determined using the generalized score matching method.  For the DPD estimator, the optimal tuning parameter is obtained as
\[
\gamma = 0.90,
\]
while for the EPD estimator, the optimal set of tuning parameters is found to be
\[
(\alpha,\beta,\gamma) = (0.1,\,0.7,\,0.1).
\]
Using these optimal values, the neural network parameters are estimated for each candidate architecture under the three estimation frameworks, namely MLE, DPDE, and EPDE.

Subsequently, for each architecture in $\mathcal{A}$, the corresponding information criteria---MLIC, DPDIC, and EPDIC---are computed based on the fitted models.  The candidate architectures are then ranked in ascending order according to the magnitude of each criterion.  From the three ranked lists obtained under MLIC, DPDIC, and EPDIC, the top candidate architectures are extracted and consolidated into a combined set.

To identify the architectures that receive the most consistent support across the three criteria, we compute the frequency with which each candidate architecture appears among the top-ranked models.  Based on this frequency, a final set of leading neural network architectures is selected. \textcolor{blue}{\hyperref[tab4]{Table~\ref{tab4}}} presents these architectures together with their selection frequencies and the corresponding values of EPDIC, DPDIC, and MLIC.

\begin{table}[H]
\centering
\fontsize{9}{11}\selectfont
\caption{Top neural network architectures based on consolidated information-criterion rankings.}
\label{tab4}
\vspace{-5pt}
\begin{tabular}{l c c r r r}
\hline
\textbf{Architecture} & \textbf{Freq.} & \textbf{Sel. Freq.} & \textbf{EPDIC} & \textbf{DPDIC} & \textbf{MLIC} \\
\hline
A4 (3,3) & 3 & 1.000 & 98853.29 & 330.1793 & 319.0178 \\
A1 (2)   & 2 & 0.667 & 113136.60 & 330.9141 & 180.5058 \\
A2 (3)   & 2 & 0.667 & 114616.58 & 331.4005 & 206.7497 \\
A3 (2,2) & 2 & 0.667 & 104669.26 & 330.2590 & 811.9595 \\
\hline
\end{tabular}
\end{table}

The results indicate that several neural network architectures receive consistent support across multiple information criteria.  In particular, the architecture with two hidden layers and three neurons in each layer, denoted by $A4 (3,3)$, appears most frequently among the top-ranked models and attains the highest selection frequency.  This suggests that moderately deep architectures provide improved flexibility for capturing the nonlinear relationships present in the predictive maintenance dataset.

At the same time, simpler architectures such as $A1 (2)$ and $A2 (3)$ also appear among the leading candidates, indicating that parsimonious network structures may still achieve competitive performance under certain criteria.  The differences observed across MLIC, DPDIC, and EPDIC reflect the varying emphasis placed on model complexity and robustness by the respective information criteria.

Overall, this consolidated ranking approach provides a robust and systematic mechanism for selecting an appropriate neural network architecture while balancing predictive performance and model complexity.  The procedure ensures that the selected architecture is supported consistently across multiple divergence-based information criteria, thereby enhancing the reliability of the model-selection process in predictive maintenance applications.

\section{Conclusion}\label{sec7}
\allowdisplaybreaks

In this paper, we proposed a robust information criterion based on the Exponential-Polynomial Divergence, aimed at addressing the limitations of classical likelihood-based model selection in the presence of contamination and model misspecification.  The proposed EPDIC leverages the flexibility and robustness of the underlying divergence to provide a reliable alternative for model selection across a wide range of settings.  We established its theoretical properties and demonstrated, through influence function analysis, that the criterion exhibits desirable robustness, with controlled sensitivity to outliers.

To ensure practical applicability, we incorporated a data-driven tuning-parameter selection mechanism based on generalized score matching, thereby improving stability over existing methods.  The effectiveness of the proposed approach was validated through extensive simulation studies and real data applications, including panel data models and neural network-based prediction tasks.  The results consistently indicated that EPDIC outperforms classical and existing divergence-based criteria in terms of stability and reliability under contaminated and misspecified scenarios.

Overall, the proposed framework provides a unified and practically viable approach to robust model selection.  Future work may explore its extension to more complex high-dimensional models and other structured learning frameworks.

\section*{Data availability}

The panel dataset used in this study is the Panel Data of Individual Wages from the Panel Study of Income Dynamics (PSID), available online at the R datasets repository: \href{https://vincentarelbundock.github.io/Rdatasets/doc/plm/Wages.html}{Panel Data of Individual Wages}.

The AI4I 2020 Predictive Maintenance dataset is obtained from the UCI Machine Learning Repository and can be accessed at: \href{https://archive.ics.uci.edu/dataset/601/ai4i+2020+predictive+maintenance+dataset}{AI4I 2020 Predictive Maintenance}.

\section*{Declaration of competing interest}

The authors confirm that they have no recognized financial conflicts of interest or personal relationships that could have influenced the work presented in this paper.

\section*{Funding}

This research is not supported by any specific grant from public, commercial, or non-profit funding organizations.

\bibliographystyle{elsarticle-num}
\bibliography{ref}

@incollection{akaike1998information,
  title={Information theory and an extension of the maximum likelihood principle},
  author={Akaike, Hirotogu},
  booktitle={Selected papers of hirotugu akaike},
  pages={199--213},
  year={1998},
  publisher={Springer}
}

@article{takeuchi1976distribution,
  title={Distribution of information number statistics and criteria for adequacy of models},
  author={Takeuchi, K},
  journal={Mathematical Sciences},
  volume={153},
  pages={12--18},
  year={1976}
}

@article{matsuda2021information,
  title={Information criteria for non-normalized models},
  author={Matsuda, Takeru and Uehara, Masatoshi and Hyvarinen, Aapo},
  journal={Journal of Machine Learning Research},
  volume={22},
  number={158},
  pages={1--33},
  year={2021}
}

@article{mattheou2009model,
  title={A model selection criterion based on the BHHJ measure of divergence},
  author={Mattheou, K and Lee, S and Karagrigoriou, Alex},
  journal={Journal of Statistical Planning and Inference},
  volume={139},
  number={2},
  pages={228--235},
  year={2009},
  publisher={Elsevier}
}

@article{basu1998robust,
  title={Robust and efficient estimation by minimising a density power divergence},
  author={Basu, Ayanendranath and Harris, Ian R and Hjort, Nils L and Jones, MC},
  journal={Biometrika},
  volume={85},
  number={3},
  pages={549--559},
  year={1998},
  publisher={Oxford University Press}
}

@article{ghosh2018new,
  title={A new family of divergences originating from model adequacy tests and application to robust statistical inference},
  author={Ghosh, Abhik and Basu, Ayanendranath},
  journal={IEEE Transactions on Information Theory},
  volume={64},
  number={8},
  pages={5581--5591},
  year={2018},
  publisher={IEEE}
}

@article{ghosh2020ultrahigh,
  title={Ultrahigh-dimensional robust and efficient sparse regression using non-concave penalized density power divergence},
  author={Ghosh, Abhik and Majumdar, Subhabrata},
  journal={IEEE Transactions on Information Theory},
  volume={66},
  number={12},
  pages={7812--7827},
  year={2020},
  publisher={IEEE}
}

@article{ray2022characterizing,
  title={Characterizing the functional density power divergence class},
  author={Ray, Souvik and Pal, Subrata and Kar, Sumit Kumar and Basu, Ayanendranath},
  journal={IEEE Transactions on Information Theory},
  volume={69},
  number={2},
  pages={1141--1146},
  year={2022},
  publisher={IEEE}
}

@article{mantalos2010improved,
  title={An improved divergence information criterion for the determination of the order of an AR process},
  author={Mantalos, Panagiotis and Mattheou, Kyriacos and Karagrigoriou, Alex},
  journal={Communications in Statistics—Simulation and Computation{\textregistered}},
  volume={39},
  number={5},
  pages={865--879},
  year={2010},
  publisher={Taylor \& Francis}
}

@article{singh2021robust,
  title={Robust inference using the exponential-polynomial divergence},
  author={Singh, Pushpinder and Mandal, Abhijit and Basu, Ayanendranath},
  journal={Journal of Statistical Theory and Practice},
  volume={15},
  number={2},
  pages={29},
  year={2021},
  publisher={Springer}
}

@article{kim2024robust,
  title={Robust estimation for general integer-valued autoregressive models based on the exponential-polynomial divergence},
  author={Kim, Byungsoo and Lee, Sangyeol},
  journal={Journal of Statistical Computation and Simulation},
  volume={94},
  number={6},
  pages={1300--1316},
  year={2024},
  publisher={Taylor \& Francis}
}

@article{warwick2005choosing,
  title={Choosing a robustness tuning parameter},
  author={Warwick, Jane and Jones, MC},
  journal={Journal of Statistical Computation and Simulation},
  volume={75},
  number={7},
  pages={581--588},
  year={2005},
  publisher={Taylor \& Francis}
}

@article{basak2021optimal,
  title={On the ‘optimal’density power divergence tuning parameter.},
  author={Basak, Sancharee and Basu, Ayanendranath and Jones, MC},
  journal={Journal of Applied Statistics},
  volume={48},
  pages={536--556},
  year={2021},
  publisher={Taylor \& Francis}
}

@article{goswami2025inequality,
  title={Inequality Restricted Minimum Density Power Divergence Estimation in Panel Count Data},
  author={Goswami, Udita and Mondal, Shuvashree},
  journal={Applied Mathematical Modelling},
  pages={116371},
  year={2025},
  publisher={Elsevier}
}

@article{xu2025generalized,
  title={Generalized score matching},
  author={Xu, Jiazhen and Scealy, Janice L and Wood, Andrew TA and Zou, Tao},
  journal={Journal of Multivariate Analysis},
  volume={210},
  pages={105473},
  year={2025},
  publisher={Elsevier}
}

@article{hyvarinen2005estimation,
  title={Estimation of non-normalized statistical models by score matching.},
  author={Hyv{\"a}rinen, Aapo and Dayan, Peter},
  journal={Journal of Machine Learning Research},
  volume={6},
  year={2005}
}

@article{bregman1967relaxation,
  title={The relaxation method of finding the common point of convex sets and its application to the solution of problems in convex programming},
  author={Bregman, Lev M},
  journal={USSR computational mathematics and mathematical physics},
  volume={7},
  number={3},
  pages={200--217},
  year={1967},
  publisher={Elsevier}
}

@article{banerjee2005clustering,
  title={Clustering with Bregman divergences},
  author={Banerjee, Arindam and Merugu, Srujana and Dhillon, Inderjit S and Ghosh, Joydeep},
  journal={Journal of machine learning research},
  volume={6},
  number={Oct},
  pages={1705--1749},
  year={2005}
}

@article{cichocki2010families,
  title={Families of alpha-beta-and gamma-divergences: Flexible and robust measures of similarities},
  author={Cichocki, Andrzej and Amari, Shun-ichi},
  journal={Entropy},
  volume={12},
  number={6},
  pages={1532--1568},
  year={2010},
  publisher={MDPI}
}

@article{nielsen2013chi,
  title={On the chi square and higher-order chi distances for approximating f-divergences},
  author={Nielsen, Frank and Nock, Richard},
  journal={IEEE Signal Processing Letters},
  volume={21},
  number={1},
  pages={10--13},
  year={2013},
  publisher={IEEE}
}

@article{kullback1951information,
  title={On information and sufficiency},
  author={Kullback, Solomon and Leibler, Richard A},
  journal={The annals of mathematical statistics},
  volume={22},
  number={1},
  pages={79--86},
  year={1951},
  publisher={JSTOR}
}

@article{ghosh2013robust,
  title={Robust estimation for independent non-homogeneous observations using density power divergence with applications to linear regression.},
  author={Ghosh, Abhik and Basu, Ayanendranath},
  journal={Electronic Journal of Statistics},
  year={2013},
  volume={7},
  pages={2420--2456},
  publisher={Institute of Mathematical Statistics and Bernoulli Society},
  doi={10.1214/13-EJS847}
}

@inproceedings{dinh2010local,
  title={Local convergence of sequential convex programming for nonconvex optimization},
  author={Dinh, Quoc Tran and Diehl, Moritz},
  booktitle={Recent Advances in Optimization and its Applications in Engineering: The 14th Belgian-French-German Conference on Optimization},
  pages={93--102},
  year={2010},
  organization={Springer}
}

@article{rousseeuw1999fast,
  title={A fast algorithm for the minimum covariance determinant estimator},
  author={Rousseeuw, Peter J and Driessen, Katrien Van},
  journal={Technometrics},
  volume={41},
  number={3},
  pages={212--223},
  year={1999},
  publisher={Taylor \& Francis}
}

@InProceedings{hubert2003pca,
    author={Hubert, M. and Rousseeuw, P. J. and Verboven, S.},
    editor={Dutter, Rudolf and Filzmoser, Peter and Gather, Ursula and Rousseeuw, Peter J.},
    title={Robust PCA for High-dimensional Data},
    booktitle={Developments in Robust Statistics},
    year={2003},
    publisher={Physica-Verlag HD},
    address={Heidelberg},
    pages={169--179},
    isbn={978-3-642-57338-5}
}

\end{document}